\documentclass[fleqn,twoside]{article}
\usepackage[headings]{espcrc2}

\readRCS
$Id: espcrc2.tex,v 1.2 2004/02/24 11:22:11 spepping Exp $
\usepackage{graphicx}
\usepackage[figuresright]{rotating}

\newcommand{\AmS}{{\protect\the\textfont2
  A\kern-.1667em\lower.5ex\hbox{M}\kern-.125emS}}

\title{As an Introduction: Quest for New Physics
in~Photon-Photon~Interactions at~the~LHC\thanks{\it Contribution to Proceedings of the CERN workshop on High Energy Photon 
Collisions at the LHC, April 21-25, 2008.}
}

\author{
 Krzysztof Piotrzkowski\address[UCL]{Universit\'e catholique de Louvain,
    Center for Particle Physics and Phenomenology (CP3)\\
    Chemin du Cyclotron 2, 1348 Louvain-la-Neuve, Belgium}\thanks{Email: 
krzysztof.piotrzkowski@uclouvain.be}
}
                       
\runtitle{Quest for New Physics
in~$\gamma\gamma$~Interactions at~the~LHC}
\runauthor{K. Piotrzkowski}

\begin{document}

\begin{abstract}
A significant fraction of $pp$ collisions at the \textsc{lhc} will involve (quasi-real) photon interactions occurring at
energies well beyond the electroweak energy scale. Hence, the \textsc{lhc} can to some extend be considered 
as a high-energy photon-photon or photon-proton collider. This offers a unique possibility for novel and complementary 
research where the available effective luminosity is small, relative to parton-parton interactions, but it is compensated 
by better known initial conditions and usually simpler final states. 
This is in a way a method for approaching some of the issues to be addressed by the future 
lepton collider. Such studies of photon interactions are possible at the \textsc{lhc}, thanks to the striking experimental signatures 
of events involving photon exchanges, in particular the presence of very forward scattered protons.
\vspace{1pc}
\end{abstract}

\maketitle
\renewcommand{\arraystretch}{1.3}
%
Photon-induced processes, and in particular the two-photon production, have not been so far a subject of intense research at proton colliders. 
In 1973 the measurement of two-photon exclusive production of lepton pairs, $pp\rightarrow pl^+l^-p$, at the 
CERN Intersecting Storage Ring was proposed as means of a precise, absolute luminosity measurement \cite{isr}. It did not materialize then, 
since at the  ISR energies it required detection of leptons at very small transverse momenta to obtain significant event statistics. Eventually, 
first such events have been observed at much higher energies, in $p\bar{p}$ collisions at Tevatron \cite{cdf} and in ion-ion collisions
at RHIC~\cite{star,phnx}, with a little background 
and very striking signature -- only two opposite charge tracks, identified as muons or electrons, without any other activity in the 
central detectors. Two tracks are back-to-back to an unusual degree, and have very little total transverse momentum. These two features 
are very characteristic of the two-photon production, due to the very small photon virtualities involved. On the one hand, 
it results in very forward scattering of incident protons, almost at zero-degree angle, on the other hand the dilepton system is produced by 
almost exact head-on collisions of quasi-real photons. Recently, the dilepton exclusive production in $pp$ collisions has been proposed 
to measure absolute LHC luminosity \cite{shamov,kp,valery}. This has been discussed also in the context of ion collisions at the LHC \cite{bocian}, 
as well as of new dedicated detectors capable to measure leptons at small transverse momenta \cite{krasny}. Corrections due to the 
strong interactions have been studied in detail and were found to be negligible, therefore the QED calculation of the cross section 
should give sufficient accuracy \cite{valery}. These features make this process a perfect calibration candle for the two-photon processes 
at the LHC, and the CDF measurement can be regarded as a proof-of-principle experiment for photon physics at hadron colliders.

When the invariant mass of the system $X$ produced in the two photon process, $pp\rightarrow pXp$ (see Fig. 1), is not too small, one can factorize 
the amplitudes for the two photon exchanges and for the photon interaction, $\gamma\gamma\rightarrow X$. This allows for introducing equivalent 
photon fluxes, which play similar role to the parton density functions for the hadron interactions. This is the basis of the Equivalent 
Photon Approximation (EPA), which allows the calculation of the proton cross-section as a product of the photon cross-section and two equivalent
photon fluxes \cite{epa}. Thanks to the huge center-of-mass energy at the LHC, $\sqrt{s}=14$~TeV the photon fluxes are significant at large
energies, well above 100~GeV, and drop with energy approximately like $1/x$, where $x=E_\gamma/E_p$ is the fraction of the incident 
proton energy carried by the exchanged photon.
\begin{figure}[htb]
\vspace{9pt}
  \includegraphics[width=16pc]{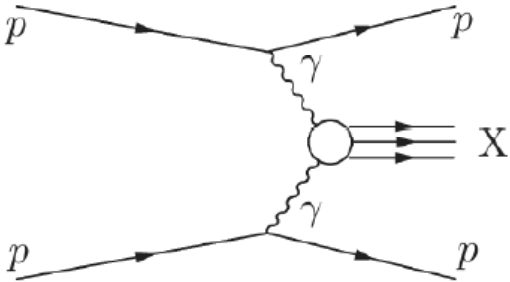}
\caption{Generic diagram for the two-photon exclusive production, $pp\rightarrow pXp$, at the LHC. Two incoming protons are scattered
quasi-elastically at very small angles, and the produced system $X$ can be detected in the central detectors.}
\end{figure}

In elastic proton scattering, when the proton survives the photon exchange, the virtuality $Q^2$ of the exchanged photon is 
limited from below by a kinematical limit, $Q^2_{min}\approx M_p^2x^2/(1-x)$, where $M_p$ is the proton mass. 
The maximal photon virtuality is then due to the presence of the proton electromagnetic form factor, and of the
finite spatial distribution of the proton charge, which cuts off completely the equivalent photon flux above 
$Q^2=1-2$~GeV$^2$.
The photon flux drops with virtuality approximately like $1/Q^2$, therefore for small $x$ the photon virtuality is on average 
very small, and one can usually treat these sub-processes as due to quasi-real photon collisions. This might be not true for inelastic
production when at least one of the two incident protons dissociates into a low mass state\footnote{One can consider also quarks as sources of the
photons, but this involves yet higher photon virtualities, resulting in much smaller equivalent fluxes. In addition, the
striking signature of lack of the proton remnants in the central detectors is then missing.}. Both minimal and maximal photon virtualities
are much higher than for the elastic case at the same energy.

The photon-photon center-of-mass energy $W_{\gamma\gamma}$ is approximately given by 
$W_{\gamma\gamma}^2=x_1x_2s$, and assuming the EPA one can introduce the two-photon luminosity spectrum in $W_{\gamma\gamma}$, 
defined as a convolution of the two photon fluxes (integrated over $Q^2$) for fixed $W_{\gamma\gamma}$.  
Such a luminosity spectrum drops approximately like $1/W_{\gamma\gamma}^2$, and its inelastic part is about  
two times bigger than the elastic one \cite{piotr}. By integrating the luminosity spectrum above some minimal 
center-of-mass energy, one can introduce the relative photon-photon luminosity $L_{\gamma\gamma}$. 
Effectively, $L_{\gamma\gamma}$ gives a fraction of the proton-proton
luminosity which is available for $\gamma\gamma$ collisions, and is especially useful if a given photon-photon
cross-section is approximately constant as a function of $W_{\gamma\gamma}$. For example, the elastic relative photon-photon 
luminosity at the LHC is equal to 1\% for $W_0=23$~GeV (i.e. for $W_{\gamma\gamma}>23$~GeV), and 0.1\% for $W_0=225$~GeV \cite{paper}.
Given the very large LHC luminosity, this leads to significant event rates of high-energy processes 
with relatively small photon-photon cross sections.  

This is even more true for the photon-proton interactions at the LHC, where both energy reach and effective luminosities 
are much higher than for the photon-photon case \cite{dis}. The high luminosity and the high c.m.s. energy of photoproduction processes offer 
interesting possibilities for the study of electroweak interaction and for searches beyond the Standard Model (SM) up to TeV scale. 
For example, one can extend studies at HERA of the single W boson photoproduction, or of the searches of anomalous Flavor Changing
Neutral Current top quark couplings,
to much higher energies \cite{paper}, effectively converting the LHC into a super-HERA collider for this class of processes.  
Some of the photoproduction cross-section are so large that become a significant part of the total inclusive cross-section at the LHC.
For example, the associated photoproduction of WH is about 5\% of the total inclusive cross-section for $pp\rightarrow WHX$ at the LHC
\cite{paper}!

\begin{figure}[htb]
\vspace{9pt}
  \includegraphics[width=18pc]{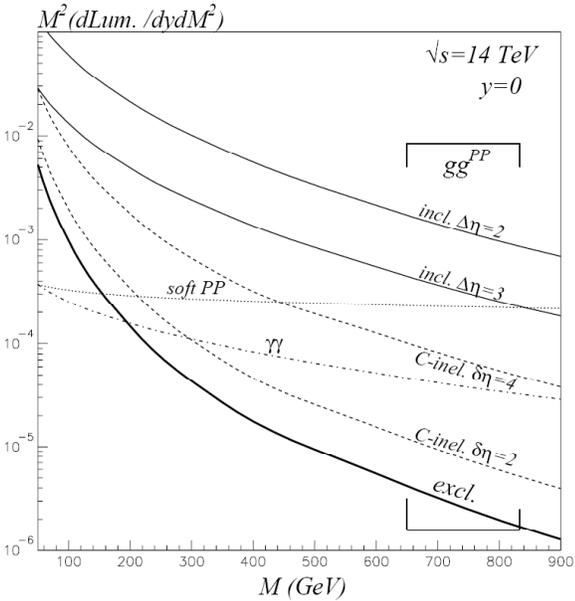}
\caption{Luminosity spectra including suppression due to the rescattering, for the exclusive photon-photon collisions 
(dashed-dotted line) and for the central exclusive diffraction (solid line) at central rapidity ($y=0$) at the LHC \cite{khoze};
$M$ is the invariant mass of the centrally produced system $X$. Other curves show the corresponding spectra for the inclusive diffraction.}
\end{figure}

One needs however to consider corrections beyond the EPA due to a possibility of strong interactions between protons, 
or the so-called rescattering effects. The resulting suppression, or the non 100\% probability of proton survival, 
of the cross sections weakly depends on the invariant mass of the exclusively produced state $X$. For the production of pairs of 
massive particles, as $pp\rightarrow pWWp$, the proton survival probability is estimated to be about 85\% \cite{khoze}. In photoproduction,
the proton survival probability is still high, though in general is slightly lower than in the photon-photon case. For example, it is 
about 70\% for the single W boson photoproduction \cite{khoze3}. In fact, the measurements of photon-induced processes were 
proposed for verification of the models describing such soft, non-perturbative strong interactions, which occur at large
impact parameter in $pp$ collisions \cite{khoze3}.

In addition, one should discuss the 
potentially dangerous background due to the central exclusive diffraction (CED). In Fig. 2, the elastic photon-photon 
luminosity is compared to the luminosity of the CED, both including the proton survival probability \cite{khoze}. The proton
survival probability is in general much smaller in diffraction due to smaller impact parameters involved, or in other
words due to smaller range of gluon exchanges with respect to the photon ones.
Plots in Fig. 2 show that for the invariant masses of the produced system above 200 GeV the $\gamma\gamma$ luminosity takes over,
so the systems exclusively produced at the LHC and of the largest invariant mass are due to photon-induced interactions. 
It means also that the CED is heavily suppressed for production of massive non strongly-interacting particles and can be safely 
neglected. For example, the gluon mediated exclusive production of W boson pairs, is about 100 times smaller than the 
two-photon production at the LHC \cite{khoze2}.

In general, the exclusive two-photon production, $pp\rightarrow pXp$, provides clean experimental conditions, and
well defined final states can be then selected, and precisely reconstructed. 
Moreover, detection of the final state protons, scattered at almost zero-degree angle, 
in the dedicated very forward detectors (VFDs), provides another striking signature, effective also at high
luminosity and with large event pile-up \cite{piotr,fp420}. A set of VFDs at 220~m or 420~m from the LHC interaction 
points will be capable of tagging photon interactions within the wide photon energy range of 
$20~\textrm{GeV}<E_{\gamma}<900~\textrm{GeV}$ \cite{hector}. In addition, the photon energies can be then well measured
with 1--5\% relative resolution and used for the event kinematics reconstruction \cite{piotr,fp420,hector}. The transverse 
momenta of the scattered protons, or equivalently the photon virtualities, are more difficult to measure, 
but providing the high resolution VFDs and precise detector alignment the absolute $p_T$ resolution of about 0.3~GeV/c could be 
achieved \cite{hector}. This then allows for some additional control of the diffractive background, which in general 
results in proton scattering at larger $p_T$ \cite{kp}. As was mentioned, the photon tagging allow to suppress very effectively, the
inclusive backgrounds, in particular it is true for double tagging of the exclusive two-photon processes, 
which is equivalent to a triple coincidence -- the detection in the same beam crossing of the system $X$ in the central 
detectors and of two protons scattered in opposite directions. This technique should be effective even at the highest
LHC luminosity \cite{fp420}. In addition, special picosecond resolution time-of-flight detectors are proposed to provide
yet another, direct control and suppression of the backgrounds due to accidental coincidences \cite{fp420}. This is based
on measuring the longitudinal position of event vertex using the {\it z-by-timing} technique, and comparing it with the event vertex z-position 
as determined by the central detectors\footnote{It should be noted that for the two-photon events this can be done using the time difference 
of the two forward scattered protons. For photoproduction it requires a very good timing for central detectors. As was shown in Ref. 
\cite{white}, for example, a 100~ps timing resolution of the central calorimeters would bring significant suppression of accidental overlay
events.}. Finally, measurements of the forward proton scattering angles provide offers also a 
unique possibility for determination of the CP parity of the produced system $X$ by measuring the azimuthal asymmetry of the 
outgoing protons, especially interesting in case of observation of new states, beyond the SM \cite{azim}. 

In the center of interest in the high energy two-photon interactions at the LHC lies, very naturally, the exclusive 
production of non strongly-interacting pairs of charged particles. It has been proposed as a possible, novel and original way
of detecting supersymmetric charginos, sleptons and the charged Higgs bosons \cite{zerwas}. But is is also an excellent
test-ground for the SM, in particular in the case of the two-photon W boson pair production \cite{tomek,saclay}.  
Detection of new, massive and quasi-stable charged particles, as predicted in some variants of supersymmetry \cite{sweet}, is very
unambiguos in two-photon exclusive production, allowing for clear interpretation. Assuming the possibility of 
measuring such events at the highest luminosity and negligible backgrounds, this type of search for new states could 
reach sensitivity to very interesting masses up to 250--300~GeV.

Finally, photon physics can be studied also in ion collisions at the LHC, where the lower ion luminosities are largely compensated
by the high photon fluxes due to the $Z^2$ enhancement (for each nucleus), where $Z$ is the ion charge.
For example, the two-photon exclusive production of the SM Higgs boson in ion collisions was considered 
as a discovery channel for a very light $H$ \cite{ihiggs}. However, for the mass of the exclusively produced system 
above 100~GeV, the coherent enhancement is not effective -- the exchanged photons prefer to couple to protons in an ion rather
than to the total ion charge $Z$. It means that at the not too high energy, where the strong electromagnetic enhancement
is still effective one can profit from an improved significantly signal-to-background ratio. 
This allows for studies of wide range of processes and new phenomena, involving in particular high parton density, nuclear 
and strong field effects \cite{ion}. Finally, it is worth adding, that in ion collisions, though in more limited
range, the proton (and/or neutron) forward tagging technique can also be applied at the LHC.

In summary, the photon-induced processes offer a rich and exciting field of research at the LHC. It shoud provide complementary
and interesting results for tests of the Standard Model as well as for search of New Physics. One cannot afford
to miss it -- on the contrary, dedicated forward detectors should be installed if one wants to get out the best of the LHC.
\section*{Acknowledgments}
The author thanks D. d'Enterria for a careful reading of the manuscript and useful comments.

\end{document}